\documentclass[12pt,preprint]{aastex}
\usepackage{pslatex}
\usepackage{verbatim}
\usepackage{graphicx}
\usepackage{amssymb}
\usepackage{lscape}
\usepackage{amsmath}
\usepackage{wrapfig}
\usepackage{float}
\usepackage{verbatim}
\usepackage{rotating}
\fontfamily{ptm}
\usepackage{hyperref}

\usepackage{pdfpages}

\usepackage{tocloft}
\usepackage{natbib}
\newenvironment{itemize*}%
  {\begin{itemize}%
    \setlength{\itemsep}{3pt}%
    \setlength{\parskip}{0pt}}%
  {\end{itemize}}

\newenvironment{enumerate*}%
  {\begin{enumerate}%
    \setlength{\itemsep}{3pt}%
    \setlength{\parskip}{0pt}}%
  {\end{enumerate}}

\newcommand{\imagedir}{./}

\usepackage{color}
\usepackage{ulem}

\begin{document}

\title{The Formation of Systems with Tightly-packed Inner Planets (STIPs) via Aerodynamic Drift}
\author{A.~C.~Boley\altaffilmark{1,}\altaffilmark{2}
E.~B.~Ford\altaffilmark{1}}

\altaffiltext{1}{Department of Astronomy, University of Florida, 211 Bryant Space Science Center, Gainesville, FL, 32611}
\altaffiltext{2}{Sagan Fellow}

\begin{abstract}
The NASA {\it Kepler} mission has revealed an abundant class of Systems with
Tightly-packed Inner Planets (STIPs). The current paradigm for planet
formation suggests that small planetesimals will quickly spiral into
the host star due to aerodynamic drag, preventing rocky planet
formation. In contrast, we find that aerodynamic drift, when acting on
an {\it ensemble} of solids, can concentrate mass at short orbital periods
in gaseous disks. Sublimation fronts may further aid this process.
{\it Kepler} data suggest that the innermost known planets are found near
the silicate sublimation zone. STIP planets should have a wide range
of volatile fractions due to aerodynamic drift and H$_2$
dissociation-driven gas accretion. We further propose that the low
mass of Mars is evidence that the Solar System was once a proto-STIP.
\end{abstract}

\maketitle

\section{INTRODUCTION}

The {\it Kepler} space mission has revealed numerous planetary types and systems that are shaping our understanding of planet formation \citep{borucki_etal_apj_2011, batalha_etal_apjs_2013}.
Among the quickly-growing data is a subclass of multi-planet configurations referred to as Systems with Tightly-packed Inner Planets (STIPs). 
Their large abundance ($>10$\% of stars) suggests that they are one of the principal outcomes of planet formation. 
The prototype STIP is Kepler-11 \citep{lissauer_etal_nature_2011,lissauer_etal_arxiv_2013}, which hosts six known transiting planets, five of which have measured masses in the super-Earth and mini-Neptune regimes. 
The known planetary orbits in this system are spaced between $a=0.09$ and 0.47 AU, with small eccentricities and mutual inclinations. 
This dynamically cold configuration suggests that gravitational interactions between the planets were minimal during planet formation or that the disk was strongly dissipative.
The orbits are not in low-order mean-motion resonances, a key signature of successive disk migration, suggesting that migration may not have played a dominant role. 
 {\it In situ} formation seems to be the simplest solution to explain the existence of STIPs. 
 However, this would require growth of massive planets on the stellar side of the ice line, which is difficult to conceive under the current planet formation paradigm. 
 We must entertain the idea that the current paradigm is incomplete and that significant massive planet formation can occur in disk regions devoid of ices.  \cite{chiang_laughlin_mnras_2013} recently explored this idea, and found that the {\it in situ} formation of super-Earths is plausible.

In this manuscript, we explore further the {\it in situ} formation of STIPs. 
 We suggest that these systems naturally arise due to rapid-followed-by-slow inward migrations of solids, draining large areas of the solid disk and building up rings. 
 We also suggest that the same process may have occurred in the Solar System, and this history is recorded in the low mass of Mars.

\subsection{THE METER BARRIER AS A PROBLEM}

The ``meter barrier'' is the seeming difficulty in gradually forming planetesimals from small solids, as aerodynamic forces will cause rapid migration of boulder-sized solids in a nearly Keplerian disk \citep{adachi_etal_1976,weidenschilling_mnras_1977}.
Because the disk has a pressure gradient, rotational and hydrostatic equilibrium for the gas will demand that the gas azimuthal velocity ($v_{ \phi}$) satisfies
\begin{equation}
\frac{1}{\rho}\frac{dp}{dr} = g_r + \frac{v_{\phi}^2}{r}\label{eqn:rohydrostatic},
\end{equation}
for radial gravitational acceleration $g_r$, gas pressure $p$,  gas density $\rho$, and disk radius $r$.  
For a monotonically decreasing pressure, $v_{ \phi}$ will always be less than the Keplerian orbital speed $v_K$.
The result is that a solid, which does not have pressure support and moves at $v_K$, will orbit with a head wind.  
This causes an exchange of angular momentum, and the solid spirals inward.  
For very large particle sizes, the stopping time ($t_s$) is  very long compared with the dynamical time at 1 AU, preventing rapid in-spiral.  
For very small particles, $t_s$ is very short, but the terminal radial velocity ($(g_r+v_{\phi}^2/r) t_s=\Delta g_r t_s$) is also very small, again preventing rapid in-spiral. 
Whenever the term $t_s v_K/r = t_s \Omega_K \sim 1$, very efficient inward drift occurs. 
Taking rough values for $r=1$ AU in an envisaged Solar Nebula ($\rho\sim 10^{-9}$ g/cc and $T\sim 300$ K), the most rapid drift size corresponds to $\sim1$ meter. 
For pressure gradients in typical disk models, the in-spiral time for a meter-sized object at 1 AU can be only a few 100 yr, much shorter than the timescale to form a planet at this location.  
This is emphasized in Figure 1, which is consistent with the results of \cite{weidenschilling_mnras_1977} (methods are explained below).  
At both smaller and larger sizes, the in-spiral time becomes long relative to 1 m.
This has been named the ``meter barrier problem,'' as it appears to inhibit planet formation.

\subsection{THE METER BARRIER AS A FEATURE}

While the meter barrier has traditionally been viewed as a problem for planet formation, it may be a critical feature. 
Aerodynamic forces will move solids, allowing them, in principle, to concentrate in select areas of disks \citep{weidenschilling_mnras_1977}.
One of the simplest solutions is to suppose that drifting solids are stopped at the inner gaseous edge of a disk,  where the azimuthal aerodynamic forces will be eliminated.
However, this may not be able to explain the wide range of planetary systems known, including the Solar System.  
If rapid inward drift occurs early in the disk's evolution, when the inner edge of the disk may only be a few stellar radii  \citep{eisner_etal_apj_2005}, the solids will collect well within the silicate sublimation region.  
If, instead, the inner hole is envisaged to be further out in the disk, as seen in transition disks \citep{espaillat_etal_apj_2012}, the disk edge can be at longer orbital periods and cooler gas temperatures.  
While appealing, there is no reason to suspect that solid evolution will wait until the transition disk phase, since significant solid evolution is already seen in young protoplanetary disks \citep{furlan_etal_apj_2009}. 
Moreover, the meteoritic record demonstrates that large solids formed very early in the Solar Nebula.
The oldest known solids in the Solar System are the Calcium-Aluminum-rich Inclusions (CAIs), which are 4568.2 Myr old \citep{bouvier_wadhwa_2010}.
Their age dispersion is very small, suggesting that CAIs formed during the earliest stages of disk evolution and set $t=0$ for planet formation.
Iron meteorite parent bodies formed over a period of about 1.5 Myr \citep{schersten_etal_2006} from CAI formation. 
Estimates of Mars's core formation suggest that Mars was half assembled by about 1.8 Myr after CAIs \citep{dauphas_pourmand_2011}.  
Taken together, this suggests that the nascent Solar System underwent significant planet formation during its earliest stages, i.e., possibly well before any transition phase, and continued for several million years.  
For these reasons, we focus on a different solution: aerodynamic terminal drift. 

Although the inward drift of meter-sized solids is rapid throughout the entirety of the inner disk, not all solids behave the same way.  
This opens the possibility that rapid relative motions of solids will induce critical solid-solid interactions.

Here we explore whether the integrated inward drift of an {\it ensemble} of particles may reach low terminal velocity drift speeds  in the inner disk  due the continuous destruction and formation of large solids.
Inward drift must not be completely stopped  as long as the ratio of the inward drift speed and the orbital speed decreases as the particle ensemble moves inward, providing sufficient time for planetoids to form.
In section {\ref{sec:results}} we present results demonstrating the basic principal of this hypothesis. 
We then outline predictions, implications, and future refinements in section \ref{sec:discussion}.

\section{SIMULATIONS\label{sec:method}}

We use a simple prescription for investigating the interactions between solids and their disk.  
We assume an idealized disk model, for which the temperature and surface density power profiles ($T=300 [r/\rm{AU}]^{-0.75}$ K and $\Sigma = 900 [r/\rm {AU}]^{-1.5}$ g cm$^{-2}$). 
The midplane density is set to $\rho_0=\Sigma/(2.5 H)$, where $H= c_a/\Omega$, for local sound speed $c_a$, and gas orbital angular speed $\Omega$. 
The vertical gas density  is calculated assuming $\rho=\rho_0\exp[-0.5(z/H)^2]$, although for the calculations here, we consider only solids that are concentrated in the midplane.
The gas orbital speed is determined using equation (\ref{eqn:rohydrostatic}).
We intend to relax these simplified approximations in subsequent studies.

Particles (solids) are integrated in cylindrical coordinates using a second-order leap frog (kick-drift-kick).
Velocities are updated during a kick step using
\begin{eqnarray}
v_r(i+1/2) & =& v_r(i) + \Delta g_r \Delta t/2\\ \nonumber
v_z(i+1/2) & = & v_z(i) + g_z \Delta t/2, \nonumber\label{eqn:kick}
\end{eqnarray}
and a drift step using
\begin{eqnarray}
\ell &=& v_\phi(i) r(i)\\ \nonumber
r(i+1) &=& r(i) + v_r(i+1/2) \Delta t\\ \nonumber
\phi(i+1/2) &=& \phi(i) + v_\phi(i)\Delta t/2 \\ \nonumber
v_\phi(i+1)& =& \ell/r(i+1)\\ \nonumber
\phi(i+1) &=&\phi(i+1/2)+v_\phi(i+1)\Delta t/2\\ \nonumber
z(i+1)&=&z(i)+v_z(i+1/2)\Delta t \nonumber,
\end{eqnarray}
which is followed by the second kick.
Here, $\Delta g_r = g_r + v_\phi^2/r$ is the excess radial acceleration relative to a circular orbit.
We use conservation of angular momentum during the drift step instead of explicitly including the Coriolis term during kicks.  
This provides stability in the leap frog method in cylindrical coordinates. 
The time step $\Delta t< (0.001) 2\pi/\Omega$ (the minimum is used for multiple particles), which was found to be sufficient for these calculations.

Next, we add drag to each kick step by writing the drag force as
${\bf F}_D =-({\bf v} - {\bf v_g})/t_s$, where ${\bf v}_g$ is the gas velocity and ${\bf v}$ is the particle velocity.  
In the Epstein limit, the stopping time $t_s$ for a solid of size $s$ and internal density $\rho_s$ embedded in a gaseous disk is given by
\begin{equation}
t_s\vert_{\rm E} = \frac{s \rho_s}{\rho c_a}\label{eqn:tstop},
\end{equation}
which is valid when the Knudsen number ${\rm Kn}={\rm mfp/(2s)}\gg1$.  
The mean free path (${\rm mfp}$) is approximated here as $\mu m_p/( \rho \sigma)$, for mean molecular weight $\mu$, proton mass $m_p$, and typical gas particle cross section $\sigma$, which we take to be $10^{-15}$ cm$^2$.  

For ${\rm Kn}\ll1$, the Stokes regime is valid, where $t_s$ now depends on the Reynolds number (${\rm Re}= 3 [\pi/8]^{1/2} \vert {\bf v }- {\bf v_g}\vert /[c_a Kn] $).
We use $t_s\vert_{\rm S} = \left[3 (8/\pi)^{1/2} k_d Kn/t_s\vert_{\rm E}\right]^{-1}$ and $k_d$ is a factor that depends on ${\rm Re}$ \citep[see][for the piecewise interpolation function used here]{boley_durisen_2010}.
We use the velocity difference at the beginning of the step to select the appropriate Reynolds number.

The total stopping time is then calculated for each particle by interpolating between the Epstein and Stokes regimes, as given above, using
\begin{equation}
t_s = \left( 9 {\rm Kn}^2+1\right)/\left( 9 {\rm Kn}^2/ t_s\vert_{\rm E} + 1/t_s\vert_{\rm S} \right).
\end{equation}

We are interested in both  large and small stopping times, including stopping times that are much smaller than $\Delta t$. 
We therefore modify the kick step such that
\begin{eqnarray}
v_r(i+1/2) & =& v_r(i) \exp[-0.5\Delta t/t_s] + (v_r^g + \Delta g_r t_s )(1-\exp[-0.5\Delta t/t_s] ) \\ \nonumber
v_\phi(i)^D & = & v_\phi(i)\exp[-0.5\Delta t/t_s] + v_\phi^g (1-\exp[-0.5\Delta t/t_s])\\ 
v_z(i+1/2) & = & v_z(i)\exp[-0.5\Delta t/t_s] + (v_z^g+g_z t_s )(1-\exp[-0.5\Delta t/t_s]), \nonumber
\end{eqnarray}
where $v_\phi^D$ is used to distinguish the drag-modified azimuthal velocity for use in the drift steps and the $g$ superscript represents gas velocity components. 
For  small $\Delta t/t_s$, the above equations approach equations (\ref{eqn:kick}) plus the drag term. 
For  large $\Delta t/t_s$, the solid velocity limits to the gas velocity components plus the  corresponding terminal velocities.

The true spatial evolution of particles will be coupled to the solid size distribution.  
Instead of considering any one  simulation particle to represent a physical, isolated solid, consider it to represent a cloud or swarm of solids. 
When a solid begins to drift away from the cloud, it can be destroyed through fragmentation or can accrete/be accreted by other solids in the cloud.  
In this way, cloud constituents are continuously being destroyed and regenerated.
The implication is that the cloud has a stopping time that is integrated over the solid size distribution.

At this time, we do not know the degree to which any one cloud is cohesive, nor do we know how the solid distribution within in a cloud will evolve. 
In the interests of moving forward, we take a highly simplified first-step approach in calculating  the behavior of aerodynamic drift in the cloud approximation.
We {\it assume} that any given cloud always maintains a fixed size distribution, here taken to be $dN/ds\propto s^{q}$ between a fixed $s_{\rm min}$ and $s_{\rm max}$. 
This implies that solid destruction and growth will not alter the relative abundances within any solid size bin, nor will it affect $s_{\rm min}$ or $s_{\rm max}$.
The degree to which this assumption reflects the behavior of an actual cloud requires a separate study.
However, even if our fixed-cloud approximation is not strictly correct, as long as it is indicative of the behavior of a swarm of solids in a protoplanetary disk, then the consequences of our assumption may reveal critical clues for understanding planet formation.
For this reason, we explore several possible $s_{\rm min}$ and $s_{\rm max}$, as well as several different $q$ ($-3$, $-3.5$, and $-4$) for one particular cloud.  
Unless otherwise stated, $q=-3.5$.

We take the cloud stopping time to be $t_s^c = \left( [\Sigma_i^M f_i m_i / t_{s,i} ]/ \Sigma_i^M f_i m_i \right)^{-1}$, where we approximate the size distribution of solids using $M$ discrete size bins, with a characteristic mass $m_i$, relative abundance $f_i$, and stopping time at that size $t_{s,i}$.
The fractional abundances are given by $dN/ds$.
For the calculations here, we use $M=8$, with each bin representing one decade in size.  
The mass is based on the number-weighted average for the given bin and an internal density $\rho_s=3$ g/cc for all solids. 
For example, a bin that presents solids between $1$ cm $<s<10$ cm with $q=-3.5$ has an average size ${\bar{s}}\approx1.3$ cm.

\section{RESULTS: Cloud Approximation for Aerodynamic Drift (CAAD)\label{sec:results}}

\begin{figure}[H]
\includegraphics[width=4.5in]{\imagedir 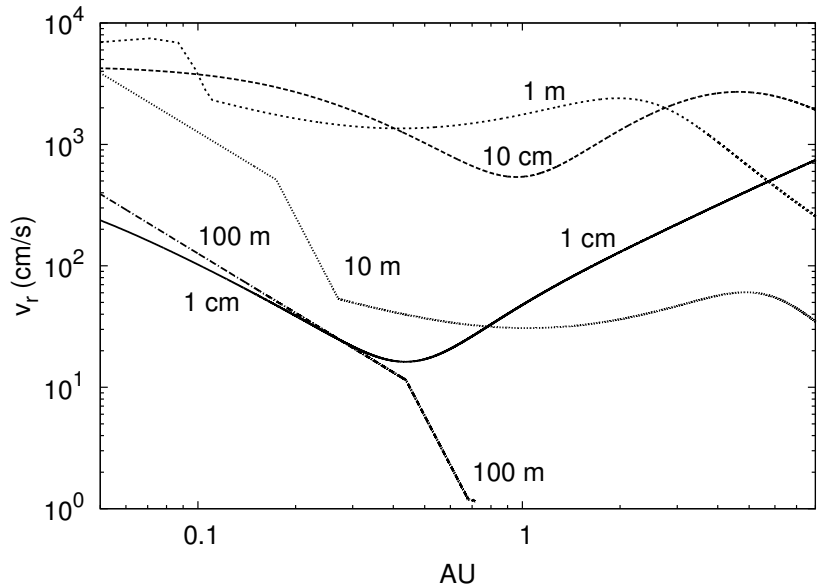}
\caption{
Illustration of the meter barrier. 
Radial drift speeds as a function of position in a disk are shown for five different solid radii, where each curve is annotated with its corresponding solid size. 
All particles except 100 m are integrated from $r=9$ AU until they reach well within 0.1 AU.  
The 100 m radius solid is initialized at $r=0.5$ AU due to its long initial radial drift time.
For 1 m-sized particles, the in-spiral timescale is only a few 100 yr.
The method for integrating the solids is given in section \ref{sec:method}.
The behavior of the solids is qualitatively similar to that found by \cite{weidenschilling_mnras_1977}.
There are quantitative differences due to the disk models used and implementations of the drag terms. 
 \label{fig:meterbarrier}}
\end{figure}

Figure \ref{fig:meterbarrier} shows the $r$-$v_r$ path of a solid of a single size $s$. The tracks are consistent with the classic \cite{weidenschilling_mnras_1977} results.
The shape of each curve reflects the different drag regimes.
Abrupt kinks for the $10$ m solid are due to the piecewise Reynolds drag solution.
While a smoother transition could be implemented, it will not change the general behavior. 
For solid sizes  $\sim 10$ cm to 10 m, the radial drift speeds become very large.
In this model, which considers an isolated, single solid size,  particles appear to be doomed to fall into the star quickly unless they encounter a pressure enhancement. 

\begin{figure}[H]
\includegraphics[width=4.5in]{\imagedir 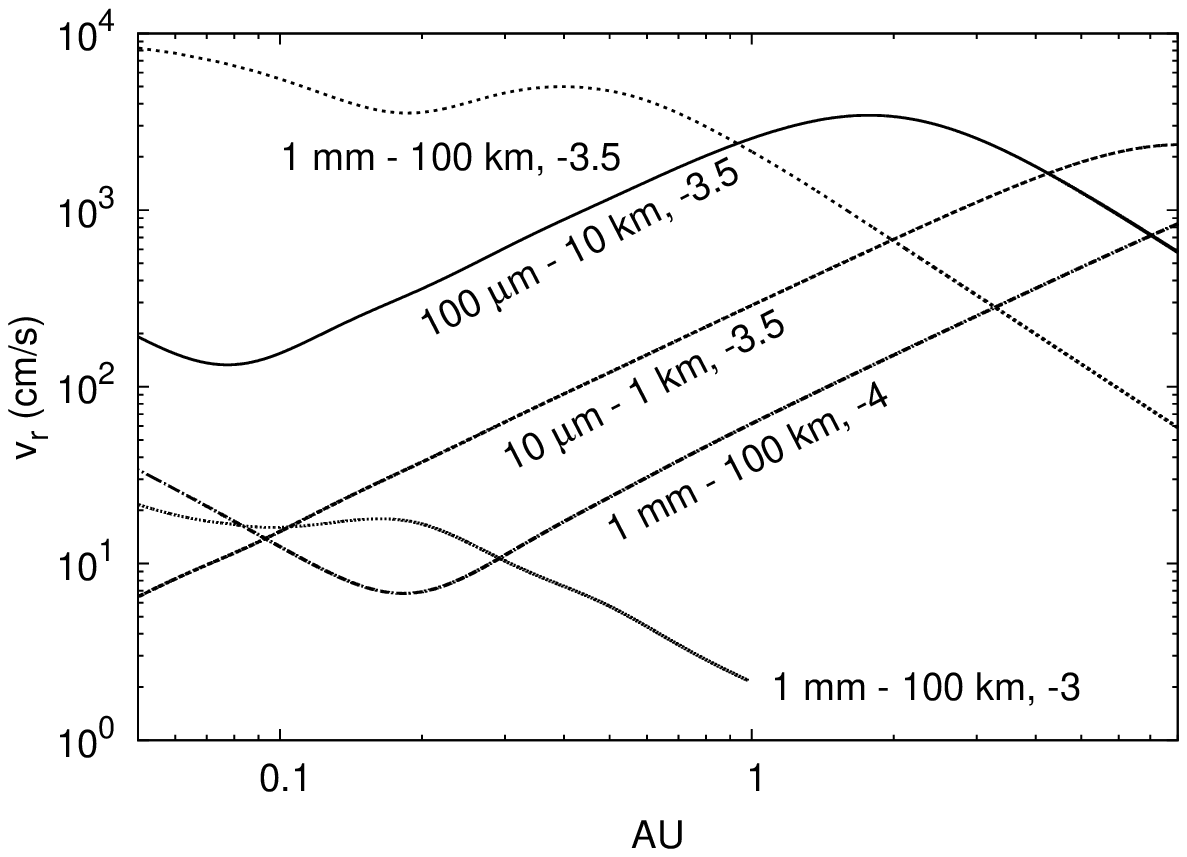}\\\includegraphics[width=4.5in]{\imagedir 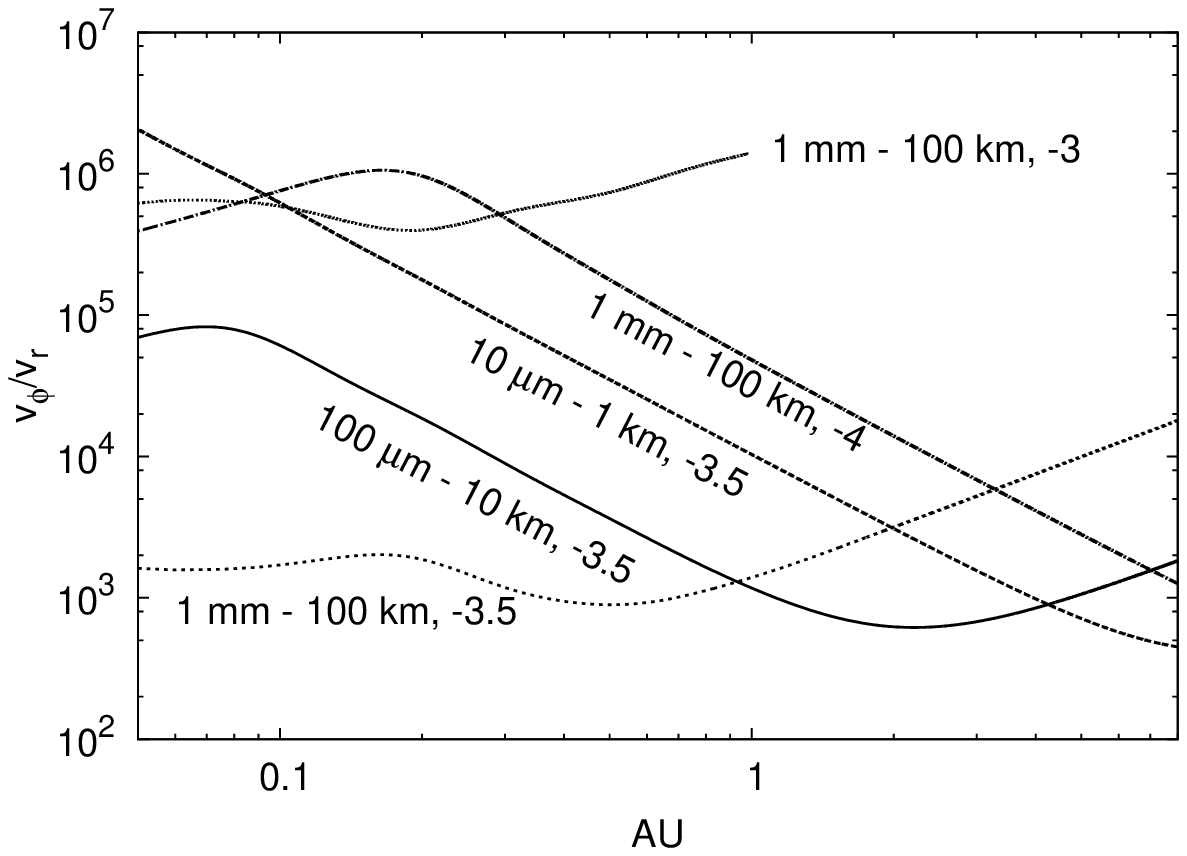}
\caption{
In-spiral using the Cloud Approximation for Aerodynamic Drift (CAAD).  
Radial drift speeds as a function of position in the same disk used for Figure \ref{fig:meterbarrier} are shown in the top panel. 
The annotations show the range of solids sizes  ($s_{\rm min}$-$s_{\rm max}$) and the power law used $dN/ds~s^{-q}$ for $q=-3$ -3.5, and -4. 
While the radial drift speeds remain high for ensembles with large $s_{\rm min}$ and $q=-3.5$, they are small compared with the azimuthal orbital speed at short periods (bottom panel).
The curves are sensitive to the assumed power law for the solid size distribution, as well as $s_{\rm min}$ and $s_{\rm max}$, but this sensitivity {\it does not change the general behavior}.
All curves show either continuously decreasing drift speeds with decreasing $r$ or have a local drift speed minimum near  $r\sim0.1$.  
For all cloud ensembles except  $1$ mm -  100 km, -3.5,  the inward drift timescale becomes $\gtrsim 10^5$ orbits at about $r=0.1$ AU,
giving significant time for the formation of planetary embryos. 
Steadily decreasing $v_r$, as well as local minima in $v_r$, may concentrate mass at small radii, triggering large planetesimal formation via gravitational instability. 
The 
 \label{fig:terminal_drift}}
\end{figure}

In contrast to the isolated solid picture, the CAAD  reveals that, if large and small solids remain coupled, then the cloud drift speed can reach small terminal drift speeds for a wide range of solid sizes and $q$ ($dN/ds\propto s^q$).  
Figure \ref{fig:terminal_drift} shows the results for five different ensembles.  
Each ensemble contains range of solid sizes ($s_{\rm min}$ -- $ s_{\rm max}$: 10 $\mu{\rm m}$ -- 1 km, 100 $\mu{\rm m}$ -- 10 km, and 1 mm -- 100 km) with either $q=-3$, $-3.5$, or $q=-4$, which is within the range of plausible $dN/ds$ distributions  \citep[e.g.,][]{pan_schlichting_2012}.
While ensembles with $s_{\rm min}\le 100~\mu{\rm m}$ have large radial drift speeds at $r\sim 1$ AU, their drift speeds steadily decrease or reach a local minimum near  $r\sim 0.1$ AU.  
The most rapid infall at short orbital periods occurs for $s_{\rm min}= 1$ mm and $q=-3.5$.  
However, changing the slope of the solid size distribution drastically changes this behavior in favor of slower radial drift speeds.  
When $q=-4$, the small solids become much more important for slowing the radial drift of the cloud.
At $q=-3$, the largest solids limit the drift.
All curves show either a steady deceleration or have a local minimum in the radial drift profile near $r\sim0.1$ AU.
This deceleration can concentrate solids to possibly high densities, allowing growth that was otherwise not possible.
In particular, the drift timescale becomes many orbital dynamical times in the inner disk, which may promote rapid solid evolution on a timescale that does not occur at larger radii.
While the cloud approximation is highly idealized, the choice of parameters for a given cloud does not seem to alter the general behavior of aerodynamic terminal drift at small disk radii.

Concentrating solids at short orbital periods may also promote gas-solid instabilities that further promote the formation of planets.
If solid ensembles converge at certain radii due to terminal drift alone, then the resulting mass build up might  initiate a streaming instability \citep{youdin_goodman_2005} in a full ring \citep{lyra_kuchner_2012}. 
Such gas-solid feedback is excluded in these simulations, and  we leave  this possibility for subsequent work.

Is the CAAD reasonable? 
A basic requirement for the cloud assumption to be valid is for the integrated collision timescale to be less than the radial drift time scale ($t_c/t_r<1$).  
In detail, this can be a complicated calculation.
For an estimate, we take $t_c\sim (\Sigma_{\rm solids} \sigma \Omega/ m)^{-1}$, where $\Sigma_{\rm solids}$ is the local surface density of solids, $\sigma$ is a representative cross section, $m$ is a representative solid mass, and $\Omega$ is the local orbital angular speed.  
This approximation yields $t_c/t_r\sim ( \langle s\rangle  \langle \rho_s \rangle /  \Sigma_{\rm solids}) (v_r/v_\phi)$, where the brackets indicate cloud-averaged properties.  
At small disk radii, $\Sigma_{\rm solid}\gtrsim100$ g cm$^{-2}$.  
From Figure 2, $v_\phi/v_r$ is much larger than 100 for all clouds.  
If $\langle s\rangle \sim 1$ cm and $\langle \rho_s \rangle \sim 1$ g/cc are appropriate for the cloud, then $t_c/t_r\ll1$ and the cloud approximation is valid, according to this criterion.

\section{DISCUSSION AND ADDITIONAL CONSIDERATIONS\label{sec:discussion}}

The results presented using CAAD suggest that significant mass can be collected in the hottest regions of the protoplanetary disk. 
This leads to multiple considerations for planet formation theory, some of which we discuss here.

\subsection{Pressure Maxima}

If any location in a disk has a local bump in the  radial pressure profile, then the pressure derivative can be negative on one side of the enhancement and positive on the other side. 
This causes regions of sub- and super-Keplerian rotation in the gas on either side, which will create head and tail winds, respectively, that will act to trap solids in the local pressure maximum \citep{adachi_etal_1976}. 
Possible causes of pressure maxima include, at least, (1) dead zones \citep{gammie_apj_1996}, i.e., areas where magnetically driven accretion (magneto-rotational instability, MRI) is suppressed, resulting in a build up of mass, and (2) evaporation fronts  \citep{kretke_lin_apjl_2007,brauer_etal_aa_2008}, where the sudden change in solids can affect accretion dynamics of the disk. 
The focus on previous work for the formation of sublimations fronts has been on the snow line, as it makes a natural connection with trapping solids to form Jupiter.  

There are multiple possible locations for pressure maxima that could lead to the formation of solid traps.
For example, the CO ice line ($\sim 25$ K) could lead to a maximum at large radii, forming a proto-Kuiper belt.
The water ice line  ($\sim 175$ K) could trap solids for the formation of moderate-period planets, corresponding to the Jovian planets in the Solar System.
Organics will sublimate at 425 K, possibly creating one of the first sublimation fronts at short periods.
Finally, around 1400 K, silicates will sublimate, setting an inner threshold for planet formation that will be at smaller disk radii for low-mass stars and at larger disk radii for high-mass stars \citep{chiang_laughlin_mnras_2013}, which we discuss next.

 \subsection{THE INNER PRESSURE MAXIMUM}
 
In section {\ref{sec:results}}, we showed that aerodynamic terminal drift of an ensemble of solids  could be sufficient for mitigating the loss of those solids to the host star while simultaneously providing the initial conditions for STIP formation.  
Nevertheless, the sudden removal of small-grain dust at the silicate sublimation zone will also affect the accretion dynamics of the protoplanetary disk.
It may be reasonable to suspect that the transition region will produce a pressure maximum that will aid terminal drift.
The pileup of inner planets within STIPs at the silicate sublimation front can be tested as more observations are acquired and theory is refined.
In the simplest model,  one could assume that the disk temperature  at a radius $r$ is set solely by radiation from the star ($T^4\approx 0.5~T_{ e}^4 [R_{\rm star}/r]^3$).
Because accretion will only make the disk hotter, this temperature represents the innermost limit for the locations of STIPs if they form and stay on the cold side of the silicate sublimation front.
Under these assumptions, the innermost semi-major axis is expected to vary with host star according to
\begin{equation}
a_{\rm in} \approx 0.8~\left(\frac{T^e_0}{1400 K}\right)^{4/3}R_0,\label{eqn:silicate_zone}
\end{equation}
where $T^e_0$ and $R_0$ are the host star's effective temperature and radius {\it during the star's PMS phase at the time of planet formation}. 
Recall that solar-type stars take $\sim10$ Myr to evolve from their birth \citep{palla_stahler_1999} to the ZAMS \citep[see, e.g.,][]{tognelli_etal_cat_2011}. 
As a result, a star's luminosity can vary substantially during the PMS phase.  
Take Kepler-32 f as an example, which is the innermost planet ($a=0.013$ AU) around a 0.54 M$_{\odot}$ star \citep{fabrycky_etal_apj_750_2012,swift_etal_apj_2013}. 
According to the Tognelli et al.~PMS tracks (for $Z=0.02$ and $Y=0.27$ $M=0.55 M_{\odot}$), when Kepler-32 was 1 Myr old, it was brighter ($L\sim1.35$ L$_{\odot}$) and larger ($\sim 2.5$ R$_{\odot}$) than the Sun is today. 
These values place $T_{\rm 1400}$ at $r\sim 0.037$ AU, well outside Kepler-32 f's current orbit.
However, if we consider the same model at 10 Myr (still on the PMS), the star is only $\sim0.14$ L$_{\odot}$  and $R\sim0.92$ R$_{\odot}$, setting $T_{\rm 1400}$ at $r=0.012$ AU, inside the planet's present orbit.

\begin{figure}[H]
\includegraphics[width=5in]{\imagedir 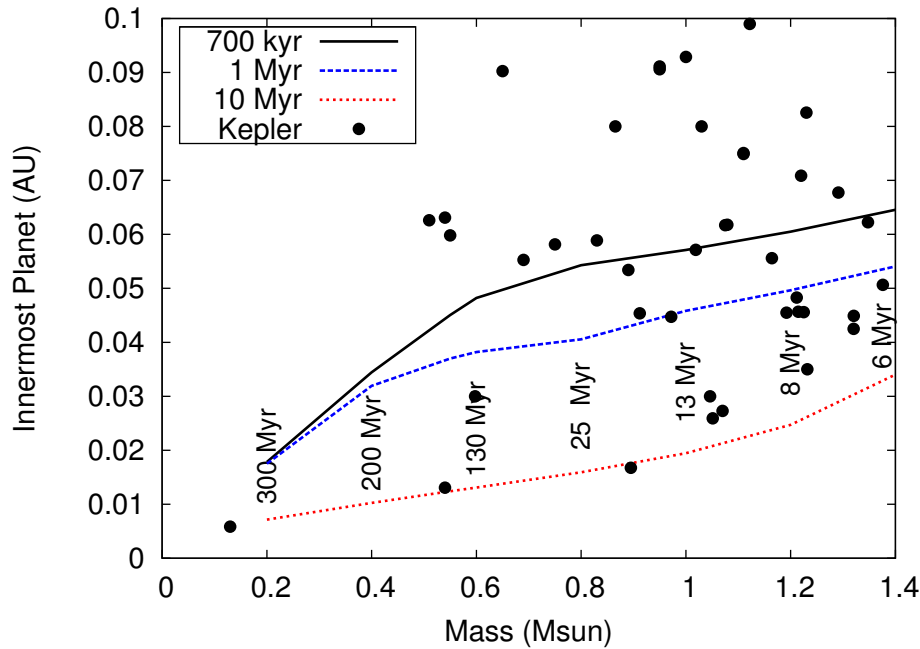}
\caption{
Possible evidence for the sublimation front acting as a barrier.
Curves show the location of $T_{\rm 1400}$ as a function of stellar mass for three different times of a star's PMS evolution. 
The black dots show the innermost planet in confirmed Kepler systems, including systems with a single planet. 
The red curve shows the absolute minimum semi-major axes one should expect for a planet based  on the luminosity and size of a given star at 10 Myr, assuming the inner location is limited by silicate sublimation. 
Annotations show approximate PMS lifetimes as a function of stellar mass.
 \label{fig:innermost_planet}}
\end{figure}

We explore the timing and inner limits of planet formation by comparing the semi-major axes of the innermost planets in confirmed {\it Kepler} systems\footnote{Data were retrieved from the exoplanets archive (http://exoplanetarchive.ipac.caltech.edu/) and exoplanets.org \citep{wright_exo}.} with their respective host star mass.
Figure \ref{fig:innermost_planet} shows the location of $T_{\rm 1400}$, as a function of host star mass, at three different times during a star's PMS evolution.  
The curves use the Tognelli et al.~evolution tables  and equation (\ref{eqn:silicate_zone}) above. 
The red curve shows the absolute minimum semi-major axes one should expect for an {\it in situ} planet based  on the luminosity and size of a given star at 10 Myr, assuming the inner location is limited by silicate sublimation. 
To date, no confirmed Kepler planet lies significantly below this curve.  
As the mass of the star increases, the planets trend to larger semi-major axis, compared with the red curve.  
If this is not a selection effect, then the trend may be due to disk heating through dissipation as the host star's mass increases above, or it may reflect the rapid evolution of the host star through its PMS phase above $\sim 1 $M$_{\odot}$ (or both). 
To illustrate the latter, the plot is annotated with approximate PMS lifetimes as a function of stellar mass\footnote{Strictly, this shows the time it takes for a PMS star to reach its first minimum in luminosity.}.

\subsection{GAS ACCRETION NEAR THE SILICATE SUBLIMATION ZONE}

 {\it In situ} gas capture near the silicate sublimation front in the inner nebula is possible \citep{chiang_laughlin_mnras_2013}, allowing the formation of gas-rich planets on short periods.
The sublimation temperature of $\sim1400$ K is interestingly close to temperatures at which molecular hydrogen gas will become highly compressible due to both vibrational modes and, at temperatures $T\gtrsim 2000$ K, dissociation. 
If H$_2$ dissociation can be initiated with only a small gravitational perturbation, then gas accretion will not be limited by radiative cooling, as energy will go into the internal gas states instead of providing pressure support.  

Consider a polytropic, non-self-gravitating atmosphere around a planet with mass $M$ and radius $R_0$. 
For the atmosphere structure relation $p=K \rho^{1+1/n}$, the density profile is
\begin{equation}
\rho_{\rm atm} = \left(\rho_0^{1/n} - \frac{GM}{K[n+1]}\left[\frac{1}{R_0}-\frac{1}{R}\right]\right)^n.
\end{equation}
We approximate the disk in which the planet is embedded to be characterized by power laws such that $\rho_{\rm disk}=10^{-9} (r/{\rm AU})^{-2.5}$ g/cc and $T_{\rm disk}=300 (r/{\rm AU})^{-0.75}$ K.
$T_{\rm disk}=1400$ K then occurs at about 0.13 AU, where the density is $\rho_{\rm disk}\approx 1.7\times10^{-7}$ g/cc.  
Taking $n=2.5$, the polytropic constant for these parameters is $K\approx 2.5\times 10^{13}$ cgs. 
The zero-density solution for the atmosphere scale height is $H= {GM}/\left( [n+1]K\rho_0^{1/n}\right)=2R_0$.
Setting $M=M_{\oplus}$ and $R=R_{\oplus}$, assuming the atmosphere remains isentropic, and letting the density at $H$ be $\rho_{\rm disk}$ instead of zero, the density at the planet's surface is $\rho_0\approx 2.3\times10^{-6}$ g/cc.  
For our polytropic model, this increases the gas temperature by almost a factor of three, well above the H$_2$ dissociation threshold at $\rho_0$ ($\sim 2500 $K).
Even the self-consistent zero-density solution is at the threshold.
Energy will go into dissociating molecular hydrogen instead of the translational degrees of freedom (pressure support). 
The result is collapse {\it without the need to radiate away significant energy}. 
In this sense, gas accretion can occur more rapidly than at larger disk radii (cooler temperatures).
The limiting mass of such an atmosphere should be explored self-consistently in future work and tested against the STIPs sample.
A sign of this process could be an enhancement of hydrogen-rich STIPs near the silicate sublimation zone, although other processes, such as hard stellar irradiation, may affect any such trend.

\subsection{CHAIN REACTIONS}

The formation of a single planet has the potential to form a pressure maximum just exterior to the newly-formed planet's orbit \citep[e.g.,][]{kobayashi_etal_apj_2012}.
The structure of resulting STIPs will be a product of initial planet masses and spacings, as well as subsequent dynamical evolution \citep[e.g.,][]{hansen_murray_2012}. 
The resulting planets need not be placed in resonances, as planet-induced pressure maxima will depend on disk properties, such as the effective viscosity \citep{tanigawa_ikoma_apj_2007}.
For the disk model used in our simulations, a core/planet that is $\sim10_{\oplus}$ would be required to open a gap at $r\sim0.1$ AU in a disk around a solar-mass star, in the no viscosity limit \citep[$H/R_{\rm Hill}<4/3$][]{crida_etal_2006}.
Whether lower planet-star mass ratios could create a chain reaction is not immediately clear.
This also needs to be explored further in subsequent work.

\subsection{IMPLICATIONS FOR COMPOSITION}

{\it In situ} formation at disk radii $\sim 0.1$ AU would seem to preclude the delivery of a substantial mass fraction of volatiles to these inner worlds.  
However, stony meteorites have been measured to contain an appreciable about of hydrated silicates, particularly in carbonaceous chondrites \citep{marty_2012,alexander_etal_sci_2012}.  
Silicates in asteroidal parent bodies that are delivered to the hot regions of protoplanetary disks through aerodynamic drag could volatilize these inner planets, even if cometary material would not survive the transport.
Depending on the role of pressure maxima, we expect a range of volatile content among STIP planets, where strong pressure maxima at organic and water ice sublimation fronts may give rise to a dry STIP, while weak pressure maxima can lead to a volatile-rich STIP.
The evolution of the host star on the PMS may also play a role in determining composition.
Because the delivery of volatiles can occur from the onset of the formation of the planets, all members of a STIP could have enrichment in volatiles, even the innermost planet.

\subsection{CONNECTION BETWEEN STIPS AND THE SOLAR SYSTEM}

We propose that the inner Solar System may have once been STIP-like due the spatial evolution of solids via aerodynamic drift. 
One large outstanding challenge for Solar System formation models is the lower mass of Mars relative to Earth and Venus.
\cite{hansen_2009} discovered that the low mass of Mars could be explained by viewing it as a stranded embryo, orbiting beyond an outer edge of a dense ring of planetesimals that form Earth and Venus.
If an initial population of planetoids are all placed within 1 AU, then subsequent oligarchic growth can lead to the formation of a Solar System-like terrestrial region, with a low-mass Mars.
While these initial conditions are admittedly {\it ad hoc}, they are nonetheless illustrative.

\cite{walsh_etal_nature_2011} suggested that Jupiter and Saturn could have created the Hansen initial conditions, in principle, through a complex migration history.
In their model (``Grand Tack''), Jupiter first migrated inward until Saturn, which migrated more rapidly, caught up to Jupiter, became locked in a resonance, and opened a mutual gap in the Solar Nebula. 
This reversed the sense of the torque, allowing the planets to migrate outward together.
If the tack occurred near 1.5 AU, an inner disk of planet embryos would be truncated to about 1 AU.

Here, we suggest that, instead of a Grand Tack among the giant planets, solids aerodynamically drifted inward, concentrated due to decreasing terminal drift speeds, and formed a series of embryos.
This could, depending on the disk and cloud properties, form  a relatively compact inner disk of solids with a depleted region near the current asteroid belt. 
Dynamical instability of the nascent inner Solar System could lead to the current spacings and the low mass of Mars, as envisaged in the \cite{hansen_2009} calculations.  
There are at least two interpretations of this scenario. 
(1) The Solar System is on the low-mass, less tightly-packed tail of a distribution of inner planetary systems.  
(2) The inner Solar System was once more massive, with a possible lost super-Earth at short orbital periods.
In either case, in this context, the Solar System owes the formation of the terrestrial planets to aerodynamic drift.
Furthermore, this suggests that STIPs are likely to have a lower-mass, exterior planet, even if not yet detected.
TTVs may reveal these planets.

\section{SUMMARY}

While individual solids near 1 meter in radius will drift rapidly into the host star, ensembles of particles, i.e., solids that are constantly being destroyed and formed, can drift slowly, with minima in $v_r$ occurring at $r<0.1$ AU for our assumed size distributions.
This terminal speed would give the solids upwards of $10^5$ orbits to evolve into planetoids, depending on the exact size distribution.
Pressure maxima at major sublimation zones will aid this process, but may not be necessary.
Terminal drift occurs  near the silicate sublimation front.  
 {\it Kepler} confirmed exoplanet data appear to have an inner limit near  $T\sim1400$K, when accounting for the PMS evolution of the star, consistent with terminal  drift-induced planet formation.
 If substantial cores form near the silicate destruction region, large atmospheres cold be accreted without the need to have efficient radiative transport, as molecular hydrogen dissociation should be initiated for relatively small perturbations by the core.
Because solids will accumulate from a range of disk radii, some silicates could be hydrated, allowing even these extremely hot planets to have an initially high volatile fraction.
We further suggest that the Solar System  may have been a proto-STIP, but dynamical instability lead to the current terrestrial configuration, including the low mass of Mars.

We thank D.~Ragozzine, G.~Laughlin, and M.~Payne for discussions that improved this manuscript.
This work was motivated, in part, through participation in the 2013 {\it Stars to Life} conference at the University of Florida.
A.C.B's support was provided under contract with the California Institute of Technology funded by NASA through the Sagan Fellowship Program. 
This material is based on work supported by the National Aeronautics and Space Administration under grants NNX08AR04G and NNX12AF73G issued
through the Kepler Participating Scientist Program.

\bibliographystyle{apj}
\bibliography{stips}


\end{document}